\documentstyle[12pt]{article}
\input epsf.tex

\begin{document}

\begin{titlepage}
\begin{center}

\hfill LBNL-38195\\
\hfill UCB-PTH-96/03\\
\hfill hep-th 9611061\\
\hfill \today\\

\vskip 0.3in

{\Large \bf \mbox{Gaugino Condensation with S-Duality and} 
\mbox{Field-Theoretical
Threshold Corrections}}\footnote{This work was supported 
by the Director, Office of Energy 
Research, Office of High Energy and Nuclear Physics, Division of High 
Energy Physics of the U.S. Department of Energy under Contract 
DE-AC03-76SF00098, and the National Science Foundation under grant
PHY-95-14797.\\ 
$\dag$ e-mail: saririan@theorm.lbl.gov }

\vskip 0.3in

Kamran Saririan$^{\dag}$

\vskip 0.2in 

{\sl Theoretical Physics Group \\
Lawrence Berkeley Laboratory \\  and \\ Department of Physics \\
University of California \\ 
Berkeley, CA 94720}\\

\end{center}

\begin{abstract}
We study gaugino condensation in the presence of 
an intermediate mass scale in the hidden sector. S-duality is 
imposed as an approximate symmetry of the effective
supergravity theory. Furthermore, we include in the K\"ahler 
potential the renormalization of the gauge coupling and the one-loop 
threshold corrections at the intermediate
scale. It is shown that confinement is indeed
achieved. Furthermore, 
a new running behaviour of the dilaton arises which
we attribute to S-duality.  We also discuss the effects of the
intermediate scale, and possible phenomenological implications of 
this model. 

\vskip 0.3in

\end{abstract}
\end{titlepage}

\newpage

\renewcommand{\thepage}{\roman{page}}
\setcounter{page}{2}
\mbox{ }

\vskip 1in

\begin{center}
{\bf Disclaimer}
\end{center}

\vskip .2in

\begin{scriptsize}
\begin{quotation}
This document was prepared as an account of work sponsored by the United
States Government. While this document is believed 
to contain correct 
 information, neither the United States Government nor any agency
thereof, nor The Regents of the University of California, nor any of their
employees, makes any warranty, express or implied, or assumes any legal
liability or responsibility for the accuracy, completeness, or usefulness
of any information, apparatus, product, or process disclosed, or represents
that its use would not infringe privately owned rights.  Reference herein
to any specific commercial products process, or service by its trade name,
trademark, manufacturer, or otherwise, does not necessarily constitute or
imply its endorsement, recommendation, or favoring by the United States
Government or any agency thereof, or The Regents of the University of
California.  The views and opinions of authors expressed herein do not
necessarily state or reflect those of the United States Government or any
agency thereof, or The Regents of the University of California.
\end{quotation}
\end{scriptsize}

\vskip 2in

\begin{center}
\begin{small}
{\it Lawrence Berkeley National Laboratory is an equal opportunity employer.}
\end{small}
\end{center}

\newpage
\renewcommand{\thepage}{\arabic{page}}
\setcounter{page}{1}
%THIS IS PAGE 1 (INSERT TEXT OF REPORT HERE)
\setcounter{footnote}{0}

{\sc 1- Introduction }

\vspace{0.2in}

A basic feature of superstring constructions in four dimensions is
 the presence of 
massless  moduli in the effective 
field theory.  These fields whose vevs
parameterize the continuously degenerate 
string vacua, are gauge-singlet 
chiral fields; furthermore, they are {\sl exact} 
flat directions of
the low energy effective field theory (LEEFT) scalar potential.  
Generically, the moduli appear in the couplings
of the LEEFT. For example, 
the tree level gauge couplings at the string scale 
depend on the dilaton, $S$, and 
the Yukawa couplings as well as the kinetic 
terms depend on the $T$-moduli (and $S$ through the K\"ahler  
potential) . There is mixing of the moduli
beyond tree level, due to both string 
threshold corrections \cite{dkl} and 
field-theoretical loop  effects.

Since the supersymmetric vacua 
of heterotic strings consist of continuously 
degenerate families (to all orders 
of perturbation theory),
parameterized by the moduli vevs, the 
latter remain perturbatively
undetermined. This degeneracy can only be lifted by a 
nonperturbative
mechanism which would induce a nontrivial superpotential for moduli,
and at the same time break supersymmetry.
We shall assume that this nonperturbative mechanism takes place in the 
LEEFT and is not intrinsically stringy. 
This certainly appears to be the most ``tractible" possibility. 
A popular candidate for such a mechanism has been gaugino condensation
which is briefly reviewed in Section 2-A. 

In  this paper, we wish to consider gaugino condensation
in a superstring-inspired effective field theory, with approximate 
S-duality invariance \cite{gz,bgsdual} and exact T-modular invariance.
We generalize the  work in ref. \cite{bgsdual} to incorporate an
intermediate scale $M_I$ ($M_{\rm  cond} \ll M_I \ll M_{\rm string}$),
and we are interested in how the intermediate-scale threshold corrections
will affect gaugino condensation and supersymmetry breaking.
The intermediate scale may be generated by spontaneous breaking
of the underlying gauge symmetry, or alternatively, by a gauge
singlet field, $A$, which is coupled to the hidden-sector gauge non-singlet
fields $\Phi_i$. In the latter scheme, 
$A$ is assumed to acquire a VEV dynamically
and therefore
gives the gauge non-singlet fields masses without 
breaking the gauge group.
We assume the latter scheme because of its simplicity. In fact, this 
scheme has been seriously considered when studying the gauge coupling 
unification \cite{unif}.
Incorporating the intermediate-scale threshold 
corrections into gaugino condensation
is non-trivial in the sense that the field-theoretical threshold
corrections at $M_I$ are dilaton-dependent. Hence, these modifications
can have non-trivial implications for supersymmetry breaking by
gaugino condensation. Furthermore, {\it a priori}, nothing 
prohibits intermediate scales in the hidden sector. 

The outline of the paper is as follows. After a brief 
review of gaugino condensation,
and of duality symmetries (modular and S-duality) in section 2, 
we shall discuss our model 
in section 3, and arrive at the the renormalized  K\"ahler  
potential including 1-loop threshold
corrections at an intermediate mass, and constrained by 
duality symmetries.  The issues related
to the scalar potential, dilaton run-away, 
and supersymmetry breaking , as well as the role of
the intermediate mass are discussed in section 3.  
Concluding remarks are given in section 4.       
\vspace{0.3in}

{\sc 2-A- Gaugino Condensation (A Review)}

\vspace{0.2in}

A possible mechanism for breaking supersymmetry within the framework
of ($N=1$, $D=4$) LEEFT of superstring is gaugino condensation in the 
hidden sector. In this scenario, the 
nonperturbative effects arise from the strong coupling  of  the 
asymptotically free gauge interactions at energies  well below $M_{Pl}$. 
Corresponding to this strong coupling is the condensation of gaugino 
bilinear  $\langle \bar{\lambda}\lambda\rangle_{h.s.}$.  
Let us briefly remind the reader the 
overview of the development of gaugino
condensation. It was recognized many years ago that gaugino condensation
in globally supersymmetric Yang-Mills theories without matter does not 
break supersymmetry \cite{vy}. In  fact, that dynamical supersymmetry
breaking cannot be  achieved in pure SYM theories was shown by
topological arguments of Witten \cite{witten}.
In the locally supersymmetric case 
the picture is rather different, namely,
gaugino condensation can break supersymmetry \cite{nilles}, 
 and the gauge coupling is itself generally 
field-dependent. When the gauge coupling
becomes strong, it gives rise to gaugino  condensation 
at the scale\footnote{These arguments are 
modified by, for instance, the 
requirement of modular invariance
\cite{bgmod}.}
\[ M_{cond}\sim M_{string}\langle 
{\rm Re}T\rangle^{-1/2}e^{-{\rm Re}S/2b_0}  =
M_{string}\langle {\rm Re}T\rangle^{-1/2}e^{-1/b_0g_{st}^2},
\] which breaks local
supersymmetry spontaneously 
 (\(M_{cond}^3 \sim\langle 
\bar{\lambda}\lambda\rangle_{h.s.} \) ),
and $S$ is the dilaton/axion chiral field.
Supersymmetry breaking in the obesrvable sector is induced
by gravitational interactions which act as `messenger'
between the two otherwise decoupled sectors. 

However, there are generally two problems 
associated with the above scenario. 
First, the destabilization of $S$ --- 
the only stable minimum of the potential in the
$S$-direction being at  $S\rightarrow \infty$; {\it i.e., } 
in the direction where exact
supersymmetry is recovered and the coupling vanishes!  
This is contrary to
the expectation that the vacuum is in the  strongly coupled, 
confining regime. This problem, the so-called dilaton runaway problem,
is present in most formulations of gaugino
condensation, in particular the 
so-called `truncated superpotential' 
approach \cite{drsw}, 
where the condensate field is assumed to be much heavier than
the dilaton and therefore is integrated out below $M_{cond}$.
In fact, the dilaton runaway problem is  perhaps a 
more generic problem 
in string phenomenology where the underlying string theory is assumed to
be weakly coupled. We shall return 
to the dilaton runaway in sections 4 and 5. 
  
The second difficulty is the large cosmological constant that arises
from the vacuum energy associated with gaugino condensation. 
An early attempt
to remedy these  difficulties was proposed by Dine {\it et al.} 
\cite{drsw}, in the
context of no-scale supergravity 
 whereby a constant term, $c\,$,  
is introduced in the superpotential which 
independently breaks supersymmetry
and cancels the cosmological constant. The origin of $c$ 
is traced to the vev of
the  3-form in 10D supergravity, and 
is quantized in units of order $M_{pl}$.
Therefore, this approach has the unsatisfactory 
feature of breaking supersymmetry at the 
scale of the fundamental theory.  
 
\vspace{0.3in}

{\sc B- Duality Symmetries}

\vspace{0.2in}

Modular symmetry, with the group $SL(2,{\cal Z})$ 
subgroup of $SL(2,{\cal R})$
 duality transformations, written in its simplest form:
 \begin{equation} T\rightarrow 
\frac{\alpha T -i\beta}{i\gamma T + \delta}, \end{equation} where
\( \alpha\delta- \beta\gamma=1\) and 
$\alpha , \beta , \gamma , 
\delta$ are integers,\footnote{There 
is, generally,  one copy of the group per 
modulus field $T^i$.} is an exact 
invariance of the underlying string theory.
However, this symmetry is 
anomalous in the LEEFT. Cancellation, or partial cancellation, of 
this anomaly in the effective theory can be achieved by 
the Green-Schwarz (GS)  
mechanism, 
which is especially clear in the 
linear-multiplet formulation of the LEEFT 
\cite{anom,anto,gt}. In the corresponding chiral formulation, the 
adding of GS counter-terms amounts to modifying  the 
dilaton K\"ahler  potential:
\[ - \ln (S+\bar{S}) \rightarrow - \ln (S+\bar{S}-bG)  ,  \]
 where $b=-\frac{2}{3}b_0$, and
$b_0$ is the $E_8$ one-loop $\beta$-function coefficient.
$G=\Sigma_{i}\ln(T^{i}+\bar{T}^{i}-
\Sigma|\Phi|^{2})$, and $\Phi$ is any
untwisted sector
 (non-modulus) chiral field in the theory. 
For simplicity, here we only
consider models where modular 
anomalies are completely cancelled by GS 
mechanism, for example, the (2,2) symmetric 
abelian orbifolds with no $N=2$
fixed planes, like $Z_3$ or $Z_7$ \cite{anom,anto,gt}.  

Recently, another type of duality symmetry has been receiving 
much attention  in string theories. In this case the 
group of duality transformations is $SL(2, {\cal Z})$, but acting on 
the field $S$ instead of $T^i$, and is referred to as 
$S$-duality. Like its $T$-analogue, this group has a generator 
which is the transformation $S\rightarrow 1/S$,
and since $S$ is related to the gauge coupling,
this duality transformation is also referred to as `strong-weak'
duality.  Font {\it et al.} \cite{font}
 have conjectured that $S$-duality,
 like $T$-duality is an exact symmetry of string theory.
More recently, there has been mounting 
evidence that S-duality is a symmetry of
certain string theories \cite{sduality}. However, these theories
all have $N=4$ or $N=2$ supersymmetries.  At the level of string theory,
there are two different types of S-duality, namely ($i$) those that 
map different theories into one another, and ($ii$) those that 
map strongly and weakly coupled regimes of the 
same theory into each another. Indeed, presently there is no evidence
of an S-dual $N=1$ string theory, and it is therefore difficult
to justify the use of S-duality as a true symmetry in the corresponding
LEEFT. However, it has been shown that in the effective 
theory, the full $SL(2,{\cal R})$ duality transformation is a 
symmetry of the equations of motion of the  gravity, gauge, and dilaton
sector in the limit of weak gauge coupling  
\cite{gz,bgsdual}. As in  \cite{bgsdual}, we shall take S-duality as a 
guiding principle in constructing the K\"ahler potential for the gaugino
condensate, which is, so far, the least understood element in the 
description of the effective theory for gaugino condensation.    
That is, we assume that S-duality invariance
is recovered in limit of vanishing gauge coupling, 
$S+\bar{S}\,\rightarrow\,\infty$.
In the Appendix , S-duality will be discussed in more detail, 
and the transformation properties of various fields under S-duality 
are given. 

There have been other recent discussions of gaugino condensation  
with S-duality \cite{snilles}
but with a rather different 
approach than ours; namely, by modifying the gauge kinetic term
by replacing the gauge kinetic function $S$ with the function 
$S + 1/S$, and introducing a 
very different nonperturbative superpotential for the dilaton than 
one gets using the
standard approach of ref. \cite{vy} 
as we do here. Other crucial differences with this work are the
renormalizartion of the dilaton in  K\"ahler  function (including 
threshold corrections), and the 
use of $SL(2, R)$ approximate symmetry (see Appendix)
to constrain $K$ in our approach.  

\vspace{0.3in}

{\sc 3- The Model }

\vspace{0.2in}

This paper basically generalizes the analysis of gaugino condensation
with S-duality of 
ref. \cite{bgsdual} to the case in the 
presence of an intermediate scale. 
Other works based on the truncated 
approach have addressed gaugino condensation
in the presence of an intermediate  scale \cite{mass}. 
However, our approach is quite different from those works in three
respects. First, the effective Lagrangian approach is adopted here rather
than the truncated approach. In the truncated approach, the mass of the 
composite is assumed to be much larger than the mass of the dilaton, 
and the condensate is  
integrated out below the condensation scale.
Here, both the composite field and the dilaton are treated as dynamical
fields. Due to this very assumption made in the truncated approach, these 
two approaches are not  equivalent in the case where the mass of the 
composite is of the order the dilaton's mass or lower. 
Second,  
invariance
under    
S-duality is used here to constrain those parts of the Lagrangian 
which cannot be obtained using the argument of anomalous symmetry. Third, the 
(dilaton dependent)
one-loop intermediate-scale threshold corrections
to the gauge coupling are included in this study.

The scheme of generating the intermediate scale considered here 
involves the 
coupling the hidden-sector 
gauge non-singlet fields $\Phi_i$ to a gauge singlet
$A$. When $A$ dynamically 
gets a vev, $\Phi_i$ become massive and the 
intermediate scale is thus generated. 
Since $A$ is a singlet, the hidden-sector 
gauge group does not break. Such a scheme has interesting 
implications for
gauge coupling  unification \cite{unif}. 
Since we are mainly interested in the 
effects of the intermediate scale 
rather than the effects of gauge symmetry
breaking, we choose the  above scheme due to its simplicity. For 
consistency, 
the pattern $M_{cond} \ll M_I \ll M_{string}$ is always assumed.
Therefore, we shall integrate out the hidden-matter fields below $M_I$ 
and the effective lagrangian at $M_{cond}$ 
will consist of the moduli and 
the gauge composites only. 

The superpotential for the hidden sector matter fields that we use is the 
following:
\begin{equation} 
W_{HM}=\frac{1}{2}\lambda^{ij}A\Phi_i\Phi_j +\frac{1}{3} \lambda'A^3 . 
\end{equation}
It is worth remarking the curious fact that in all the examples of 
semirealistic superstring models 
with exactly three generations of matter 
that have been studied so far 
\cite{shyamoli} no cubic {\sl self-coupling} 
of gauge singlets seems to arise 
in the superpotential. However, there are 
indeed cubic couplings in the 
superpotential that involve two or three 
different gauge singlets ($\kappa_{\alpha\beta}
 (A^{\alpha})^2A^{\beta}$ or 
$\kappa_{\alpha\beta\gamma}A^{\alpha}
A^{\beta}A^{\gamma}$ with $\alpha$,
$\beta$, and $\gamma$ all different). 
The  cubic self-coupling is, however, not 
ruled out on any physical grounds. 
So, just to be consistent with the current 
literature, one should perhaps introduce at 
least a pair of gauge singlets, 
one of which is coupled to the 
gauge-charged matter fields. In that sense 
our case is a toy model  
describing the situation 
where the gauge singlets
have mutual couplings comparable 
to our $\lambda'$. However, for the 
general analysis of gaugino 
condensation in the presence of an intermediate 
scale, no new feature can be 
expected  to arise from the extra gauge 
singlets as compared to our simplified case.    

When constructing our model, two symmetry 
principles have been used to
constrain the Lagrangian: 
First, the LEEFT must be T-modular invariant to
all orders. Second, S-duality
is a symmetry in the weak-coupling limit 
$\langle S+\bar{S}\rangle \rightarrow \infty$, 
as has been discussed in Sec.2-B.
Furthermore, we adopt also the point of 
view that the K\"ahler potential
is renormalized instead of the gauge kinetic term when 
including the 
renormalization effects of the tree-level gauge coupling $S+\bar{S}$.
This viewpoint is especially clear in the linear-multiplet formalism of
the LEEFT. 
For example, in the linear-multiplet formalism, the cancellation
of modular anomaly is achieved 
by adding the Green-Schwarz (GS) counterterm
through the linear multiplet, which contains the string two-index 
anti-symmetric tensor field. When going from the  linear-multiplet
formalism to the chiral formalism by performing the 
supersymmetric duality
transformation, the GS counterterm of the linear-multiplet formalism 
transforms into the renormalization 
of the tree-level coupling $S+\bar{S}$
in the K\"ahler potential of the 
chiral formalism only. The gauge kinetic
term of the corresponding chiral 
formalism remains unrenormalized \cite{gt}.
Hence, we
will include the renormalization and intermediate-scale 
threshold corrections only in the 
dilatonic part of the K\"ahler potential.
It is worth noting that   in the exact S-duality 
limit, 
in our chiral multiplet approach,
the superpotentials for the matter field as well as the
chiral condensate are absent. 
In constructing an effective theory for the chiral condensate
field,
consistent with the symmeteries of the underlying theory 
(modular and $S$ duality symmetries), we  include the 
wave function
renormalization of the condensate, $H$,  in the K\"ahler  potential.
 Put differently, the usual superpotential $W_{NP}\sim H^3$ 
is absent by requiring S-invariance in the $g\rightarrow 0$ limit; and  
so in the effective theory 
for this field, 
rather than having a quantum correction of the form $W_q\sim
H^3\ln(H/\mu)$, we have a renormalization of the K\"ahler  potential 
corresponding to  the wave function renormalization of $H$. 

Let us start with the construction of the K\"ahler potential. 
We derive the K\"ahler potential $K$ in two slightly different
ways. The first derivation is straightforward: we take the canonically 
normalized mass of the fields $\Phi_i$ (which is a field-dependent,
modular-invariant quantity) as the dynamically generated intermediate 
scale $M_I$, and the gauge coupling at the condensation scale is obtained
easily by running the gauge coupling from the string scale
first to the intermediate scale, and then to the condensation scale 
together with the fact that the matter fields 
of mass $M_I$ decouple below 
$M_I$. 

In the second derivation, we apply the result derived in ref. \cite{kl1}
for the corrections to the gauge coupling at a 
field-theoretical threshold
to one loop. Their result was derived for a generic supergravity
model with a threshold scale, with no reference to modular invariance.
In a modular invariant theory, we can show that both 
approaches result in the
same gauge coupling, and therefore the same K\"ahler potential $K$.

The no scale case of the   K\"ahler potential \cite{bgsdual} (without
matter fields, {\it i.e., } with pure $E_8$ gauge group)  
at the condensation
scale  is given by 
\begin{equation} K=-\ln m_0 -3\ln(1-m_0 ^{1/3}Q) + G \end{equation}
where,
\begin{equation} 
m_0=S+\bar{S} -bG +3b\ln Q 
;\hspace{0.2in} G=-3\ln(T+\bar{T} ); \hspace{0.2in}
Q=|H|^2e^{G/3};\hspace{0.2in} -3b=2b_0=\frac{1}{8\pi^2}C(E_8) .
 \end{equation}
Here, $M_{pl}=1$; and notice that the UV cut-off is taken
to be $M_{string}=(S+\bar{S}-bG)^{-1/2}$ meaning that the condensation
scale is really in  these units, $Q/(S+\bar{S} -bG)$.

In the presence of an intermediate scale the  renormalization of
the gauge coupling in $m_0$ will be different from that of 
ref. \cite{bgsdual}. 
If we include the threshold corrections at one loop, we  get 
\begin{equation} m_0 \rightarrow m = S+\bar{S} -b G 
% -2b^>_0\ln\left(\frac{M_I^2}{M_{string}^2}\right)
% -2b^<_0\ln\left(\frac{Q/(S+\bar{S}-bG)}{M_I^2}\right) \nonumber \\
% S+\bar{S} -bG
 +3\left[b^>\ln\left(\frac{M_I^2}{M_{string}^2}
\right) +b^<\ln\left(\frac{Q/(S+\bar{S}-bG)}{M_I^2}\right)\right] ,
\end{equation}
and the K\"ahler  potential at the condensation scale is:
\begin{equation} K=-\ln m -3\ln(1-m ^{1/3}Q) + G . \end{equation}
Here, $b^>$  and $b^<$ are proportional to the $\beta$-function coefficients 
above and below the intermediate scale, respectively:
\begin{equation} b^> = (3C_G -C_M)/24\pi^2 ,
 \hspace{0.3in} b^< = C_G/8\pi^2 , \end{equation}
where $C_G$ and $C_M$ are the quadratic Casimirs:
\begin{equation} C_G=T(adj); \hspace{0.2in} C_M=\sum_{r}n_r T(r); 
\hspace{0.2in} T(r)={\rm Tr}_r(T^2), \end{equation}
with $r$ labelling the 
representations of the gauge group,  and $n_r$ being the
number of fields in the $r$ representation. As expected, in the 
absence of $M_I$, {\it i.e., } for $b^>=b^<$,  we recover 
the K\"ahler  potential of ref
\cite{bgsdual}. Let us briefly note that the general form of
the  K\"ahler  potentials 
(3) and (6) is simply obtained by starting  with the
modular invariant tree
level K\"ahler  potential  (supplied 
with the approrpiate GS counter-term, G)
which includes
the kinetic term for the condensate field, $H$:
\[ K=-\ln(S+\bar{S}-bG)-3\ln(e^{-G/3} - f(S,\bar{S})|H|^2) , \]
and imposing S-invariance, which gives 
$f=(S+\bar{S})^{1/3}$ up to
an S-invariant factor (see Appendix). 
Finally one replaces $S+\bar{S}-bG$ with 
the one-loop renomalized effective coupling at the 
condensation scale, 
which we have denoted $m=2/g^2_{eff}(M_{cond})$.

The modular invariant scale $M_I$ has to be determined  --- it is the
modular invariant, canonically normalized mass of $\Phi$, and
not simply the vev of the gauge singlet $A$, which has a nonzero
modular weight. Before computing $M_I$, let us make the distinction  
between the GS terms above and below the threshold, namely,
\begin{eqnarray}  G^> &=& -3\ln(T+\bar{T} - |A|^2 -\sum|\Phi_i|^2),  
\nonumber \\
G^< &=& -3\ln(T+\bar{T} - |\langle  A\rangle|^2) . \end{eqnarray}
Indeed, the difference arises only due  to the change in the spectrum
as the threshold is crossed. 
`We  analyse the theory with all the massive fields ($\Phi_i$ and
$A$) ``integrated out" at the condensation scale, so that in the 
first line of eq (9), these fields are replaced with
their vacuum expectation values to obtain
$G^<$ in the second line. We discuss what kind of an approximation 
this replacement entails at the end of this section. 
  
It is straight forward to show that the canonically normalized mass
is:
\begin{equation} 
M_I^2=e^K(K^{\varphi\bar{\varphi}})^2|\lambda A|^2=
\frac{|\lambda A|^2e^{G/3}}{9(s+\bar{s}-bG)}
\left(1 + \frac{b}{s+\bar{s}-b G}\right)^{-2} . \end{equation}
Modular invariance is automatic due to the appropriate G-S terms,
provided that $A$ has the following modular transformation property:
\(A\rightarrow|i\gamma T + \delta|^{-1}A\), {\it i.e., }  
has modular weight $-1$.

We now derive the above K\"ahler  potential by a different argument.
It can be shown \cite{kl1} that in the presence of
a mass, in a YM + supergravity effective theory, the gauge couplings
receive threshold correction 
at a scale $\Lambda_I$ given by
\begin{eqnarray} \frac{1}{g^2(p_<)}-\frac{1}{g^2(p_>)}
&=& b^>_0 \ln \frac{p_> ^2}{\Lambda_I ^2}  
+b^<_0 \ln \frac{\Lambda_I ^2}{p_< ^2} 
-(c^> - c^<)K \nonumber \\
&-& \frac{1}{8\pi^2}(T(adj)^>-T(adj)^<)\ln g^{-2}
+ \frac{1}{8\pi^2}\sum_r T(r)\ln\det Z_{\rm massive}^r . \end{eqnarray}
The group
theoretic factors $c^>$ and $c^<$ are respectively given by:
\begin{equation} c^>= (-C_G+C_M)/16\pi^2 ; \hspace{0.3in} c^< =-C_G/16\pi^2  ; 
\hspace{0.3in} c^>-c^<=b^>_0-b^<_0=-3/2(b^>-b^<)=C_M/16\pi^2 . \end{equation}
 The K\"ahler  function, wave function renormalization matrix $Z$ of the
massive fields, and the (effective) coupling $g$ on the right hand side of the 
above formula are all {\sl tree level} quantities at the intermediate
scale.  The  derivation 
of the above equation assumes 
noncanonical normalization of the tree level kinetic terms in the 
supergravity Lagrangian. In particular, modular invariance 
plays no role, and the intermediate scale is not fixed by 
modular invariance and canonical normalization.   Therefore, we take  
 $\Lambda_I ^2=|\lambda A|^2$ as one would in the noncanonical 
normalization. 
The K\"ahler  term in eq. (11)
must contain the contribution of the massive fields, {\it i.e., } it is
$K_{tree} = -\ln (S+\bar{S}-b G) + G$, with $G$ given in the first line
of eq. (9).
For the UV cut-off, we use $M_{string}$ and
for condensation scale $M_{cond}^2=Q/(S +\bar{S} -bG)$, as before.
The normalization matrix for the $\Phi$ fields is given simply by
the K\"ahler  metric components 
$K^>_{I \bar{J}}$.  One only finds contributions
from the diagonal blocks:
\( Z_{I\bar{I}}=3[1+b/(S+\bar{S}-b G)]e^{G/3} .\) 
Hence, 
\begin{eqnarray}  2\sum_r T(r)\ln\det 
Z_{\rm massive}^r &=& 2\sum_r T(r)\sum_I
\ln\left[3e^{G/3}\left(1+\frac{b}{S+\bar{S}-b 
G}\right)\right]_r \nonumber \\
%&=& \sum_r T(r)n_r \ln \left[3\left(1+
%\frac{b}{S+\bar{S}-b G}\right)e^{G/3}
%\right]^2 
%\nonumber \\
&=& (b^>_0-b^<_0)\ln 
\left[3\left(1+\frac{b}{S+\bar{S}-b G})e^{G/3}\right)
\right]^2 ,
\end{eqnarray}
Finally, notice that in our scheme of generating the 
intermediate mass the $T(adj)^>-T(adj)^< =0$, and thus the 
corresponding term in the threshold correction will be absent.
Making the above replacements in eq. (11) gives:
\begin{eqnarray} S+\bar{S}-b G \rightarrow  
% S+\bar{S}-b G +2\left[ -b^>_0
% \ln \left(\frac{|\lambda A|^2}{M_{string}^2}\right) -b^<_0 
% \ln \left(\frac{Q/(S+\bar{S}-b G)}{|\lambda A|^2}\right) \right]
% \nonumber \\
% &-& (b^>_0 -b^<_0)\left[G- \ln(S +\bar{S} -bG)
% -\ln \left( \frac{e^{G/3}}{9(1+\frac{b}{S+\bar{S}-b G})^2}\right) 
% \right]
%  \nonumber \\ 
 S+\bar{S}-b G &+& 3 b^> \ln \left(
 \frac{|\lambda A|^2
e^{G/3}(1+\frac{b}{S+\bar{S}-b G})^{-2}}{9(S+\bar{S}-b 
G)M_{string}^2} \right) 
\nonumber \\
&+& 3 b^< \ln \left( \frac{Q}{|\lambda A|^2
e^{G/3}(1+\frac{b}{S+\bar{S}-b G})^{-2}/9} \right)  , \end{eqnarray}
which is precisely the same as in eq. (5). 
%\begin{eqnarray}
% S+\bar{S}-b G & \rightarrow & S+\bar{S}-b G + 3\left[
% b^> \ln \frac{M_I^2}{M_{string}^2} + b^< \ln
%\frac{Q/(S+\bar{S}-b G)}{M_I^2} \right] \nonumber  \\ 
%& = & S+\bar{S}-b G + 3\left[
% b^> \ln \widetilde{M}_I^2 + b^< \ln
%\frac{Q}{\widetilde{M}_I^2}, \right]
%\end{eqnarray}
%Where $\widetilde{M}_I^2=M_I^2 g_{string}=M_I^2(S+\bar{S}-b G)^{-1}$. 
%This is in exact agreement with what we had in  eq. (5).
 To summarize, our K\"ahler  potential
is given by eq. (6) and (5) which is the extension 
of that proposed in ref. 
\cite{bgsdual}.
 This 
extension consisted of 
the renormalization of the gauge coupling in $K$, including the
one-loop field-theoretical threshold 
corrections around $M_I$, the modular
invariant, canonically normalized intermediate mass scale. 
%  
% \vspace{0.3in}%%
%
% {\sc 4- Integrating Out the Heavy Fields}
%
%\vspace{0.3in}

A comment on integrating out the 
heavy fields and replacing them with 
their vevs is perhaps in order. 
We have obtained the renormalized 
K\"ahler  function at the condensation scale. 
Since the masses of the heavy 
fields (${\cal O}(M_I)$) are, by assumption,  
much  larger than the condensation 
scale, we must integrate out all the 
heavy fields. We assume that  the gauge-singlet $A$ is heavy,
with $M_A \sim {\cal O}(M_I)
 \gg M_{cond}$; {\it i.e., } the self coupling of 
$A$ in the superpotential 
$W(A)=\frac{\lambda'}{3}A^3$ is sufficiently large.
Then it is easy to show that 
if we integrate out the fields $A$ and $\Phi^i$
at tree level, the following terms are 
generated in the effective potential: 
\begin{eqnarray} 
V_{eff}&=&M_A ^{-2} K^{a\bar{a}}|
\left[ (K_{i\bar{\jmath} a}K_{\ell\bar{m}\bar{a}})|
\partial_{\mu} 
z^i\partial^{\mu}\bar{z}^{\bar{\jmath}}\partial_{\nu}z^\ell 
\partial^{\nu}
\bar{z}^{\bar{m}} +(V_aV_{\bar{a}})|  
  -  \left( (V_a K_{i\bar{\jmath}\bar{a}})|\partial_
{\mu}z^i\partial^{\mu}\bar{z}^{\bar{\jmath}} + h.c. \right)
\right] \nonumber \\
 &=&(M_A^{-2}K^{a\bar{a}}K_{i\bar{\jmath} a}K_{\ell\bar{m}\bar{a}})| 
\partial_{\mu}
z^i\partial^{\mu}\bar{z}^{\bar{\jmath}}\partial_{\nu}z^{\ell} 
\partial^{\nu}
\bar{z}^{\bar{m}}.
\end{eqnarray}
The quantities denoted by a vertical bar are evaluated at the 
vacuum ($a=\langle a \rangle$ , $\varphi^i=\langle \varphi^i \rangle=0$).
The last line follows from the fact that
  $V_a=\partial V/\partial a$ vanishes at $\langle A \rangle$. 
Since, the effective potential (20) that arises  
contains  only 4-derivative couplings,
 at energies well below $M_A$, {\it i.e., } at the condensation
scale it can be ignored, and in our analysis, 
we can replace the heavy fields with their vev's.

We close this with the following remarks. We notice that 
a constant term is 
generated in the superpotential, namely
\begin{equation}   c = \frac{\lambda'}{3}\langle A \rangle^3. 
\end{equation}
In essentially all models of gaugino condensation, 
introduction of a constant 
superpotential is necessary for breaking
 supersymmetry. However, the constant is usually
either introduced in an {\it ad hoc} way, or its 
origin is from compactification of 
superstrings. Namely, the vev of the compactified
components of the 3-form, $H_{lmn}$ from 10-D
supergravity \cite{drsw}.
 In the latter case, the constant has
the undesirable property that  it is of the 
order of Planck mass (thus breaking supersymmetry at $M_{Pl}$)
and that it is quantized, presumably in units of $M_{Pl}$.
The above constant $c$ is clearly much smaller (of the order 
of $M_I$ and it is continuous.
The second remark has to do with the fact that we know (see eq. (9)) 
that $|\langle A\rangle|^2 < \langle T+\bar{T}\rangle$.
Further, we know that the vev of $T$ is not determined 
perturbatively. The nonperturbative superpotential
for the condensate is what will eventually allow us to 
fix  $\langle T\rangle$. So, how are we
justified in
integrating out $A$ but not $T$? The only justification we
offer  is the  fact that the $T$ modulus remains massless 
to all orders in perturbation theory until supersymmetry is broken 
(nonperturbatively)
by the gaugino  condensation (or otherwise), whereas $A$ is by construction
massive ($M_A\sim M_I$).    
    
\vspace{0.3in}

{\sc 4- Scalar Potential and the Vacuum}

\vspace{0.3in}

The dynamical fields at the condensation scale in our
model are $S$, $H$, and $T$. 
The scalar potential is given by:
\begin{equation} V=e^K\left[ K^{i\bar{\jmath}}(K_iW+W_i)
(K_{\bar{\jmath}}\bar{W}+\bar{W}_{\bar{\jmath}}) - 3|W|^2\right],
\end{equation}
and the K\"ahler  
metric written in terms of $m=2/g_{eff}^2(M_{cond})$ (eq. (5)),
$Q=|H|^2e^{G/3}$, and their derivatives 
with respect to the scalar fields
is given by:
\begin{eqnarray}
K_{i\bar{\jmath}} = m^{-2}\{ m_im_{\bar{\jmath}}\tilde{x} 
&+& m(\xi-1)m_{i\bar{\jmath}} + (\xi+\xi^2)
(m_iq_{\bar{\jmath}}+m_{\bar{\jmath}}q_i) \nonumber \\
  &+& 3m^2[\xi q_{i\bar{\jmath}}+
(\xi+\xi^2)q_iq_{\bar{\jmath}}]  + m^2 G_{i\bar{\jmath}} \}  ,
\end{eqnarray}
where 
\[x=m^{1/3}Q, \hspace{0.2in} \xi=\frac{x}{1-x}, \hspace{0.2in}
\tilde{x}=1-2\xi/3+\xi^2/3 ,\] and
\[m_i=\partial_i m , 
\hspace{0.2in} q= \ln Q ,\hspace{0.2in} q_i=\partial_i\ln Q
, \hspace{0.2in} etc. \]
Notice that $G_{i\bar{\jmath}}=0$ unless $i=j=t$,  
$m_{h\bar{\jmath}}=0$, and $q_s=0$.  
The  nonperturbative part of the superpotential is of the form
\begin{equation} W_{NP} = \alpha e^{-S/b}Y^n \left(\ln \frac{Y}{\mu}
\right)^k , 
\hspace{0.2in} Y=He^{S/3b^<}\end{equation}
with  $n<3$ (the Veneziano-Yankielowicz superpotential 
is the special case of 
$n=3$ and $k=1$). The reason the exponents $n$ and $k$ are introduced
is because, as stressed earlier, it is the 
 K\"ahler  potential (6) that already includes the gaugino condensate
 wave function renormalization,
and so the superpotential should not.   

Is the potential positive semi-definite? Numerical analysis
  indicates that
the answer is yes. Analytically, this would be obvious if 
$\langle W \rangle$ could be 
shown to be zero. In fact, numerically\footnote{In
the numerical analysis, the value of $\langle $Re$t \rangle$ 
was fixed and $s$ and $h$
were  varied (see later).} we find that at the minimum of $V$,

\noindent $\bullet$ \(\langle W \rangle\simeq 0\).

\noindent $\bullet$ \(m=2/g_{eff}^2(M_{cond})\rightarrow 0\).

\begin{figure}
\begin{center}
\leavevmode
\epsfysize=3in \epsfbox{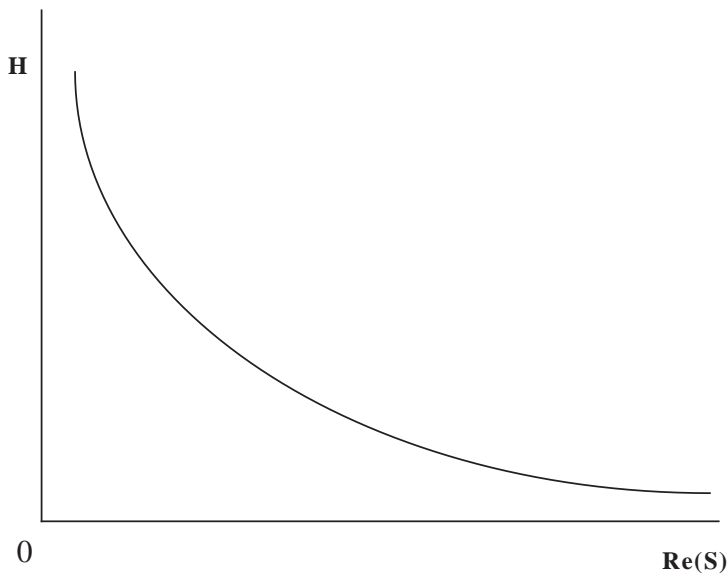}
\caption{The boundary between the kinematically forbidden (below the
curve) and allowed regimes contains the nontrivial minimum of the 
scalar potential $V(s,h)$.} 
\end{center}
\end{figure}

\noindent To see  that $\langle W \rangle=0$ at $V=0$, guided by the 
second numerical result above, we expand $V$ in 
powers of $m$ in the limit \(m\rightarrow 0\).
 A lengthy but straightforward calculation shows that 
when $\langle W \rangle=0$,
\(V\sim {\cal O}(m) +\) higher order as \(m\rightarrow 0\); 
and for \(\langle W \rangle\neq 0\),
there would be a pole $\sim 1/m$ in $V$ (this is because 
the threshold corrections at $M_I$ cause the K\"ahler  potential
to be no longer exactly `no-scale').  No such pole was found; the minimum
of $V(s,h)$ corresponds to the minimum of $m(s,h)$ (which is zero). The
analyical asymptotic expansion of $V$ in $m$, and the numerical results 
are compatible only for $\langle W \rangle=0$.   
The reality of the K\"ahler  
function, and the hierarchy \(M_{cond}\ll M_I\ll
M_{pl}\) restrict the kinematically 
allowed region of the parameter space
such that:\footnote{Hereafter, lower case letters indicate the scalar
components of the 
corresponding superfields.}
\[ a<\sqrt{2t}, \hspace{0.2in} \lambda a/3 \;\gg  h \]
(for simplicity, we take both $s$  and $h$ to be real). 
The kinematically
forbidden and allowed regions are typically separated as shown in
Fig. 1.  The boundary between the two regions contains the nontrivial 
minimum  satisfying $m=0$ and 
$\langle W \rangle=0$ (as well as the trivial minimum
$(s,h)=(\infty,0)$). 

Both  $m$ and $\langle W \rangle$ increase 
monotonically from zero in both $h$ and Re$ s$ 
near the vacuum\footnote{`Vacuum' here 
refers to the nontrivial minimum of the 
potential.}.
The plot shown in Fig. 2 shows that $V(s,h)$ also 
monotonically increases in both directions, and particularly sharply 
in the $h$ (condensate) direction, indicating confinement.
 In the direction of the 
dilaton, the potential increases quadratically as
a function of $S$. This 
can be seen by looking at the  $S$-dependence of 
 $V(m\approx 0)$. Furthermore, we notice that
the dilaton does `run away', but in the correct direction! Namely,
to some finite value of $s$ (which 
separates the kinematically allowed and forbidden
regions, at the nontrivial minimum of the potential). This is in 
addition to the usual runaway behaviour to $s\rightarrow \infty$, which
is the susy-restoring and deconfining limit. Also interesting 
 is the behaviour of $m=2/g_{eff}^2(M_{cond})$ near the 
vacuum, which as
noted above, is \(m\rightarrow 0\) or \(  g_{eff}(M_{cond}) \rightarrow
\infty\) (while $g_{st}$ remains finite).
 This is exactly what one expects physically, since the condensate --- 
the  bound 
state in the strong coupling regime ---
  is expected to correspond  to a stable vacuum
solution. Notice, however, that the relations 
$\langle  V \rangle = \langle  W \rangle = 0$
imply that supersymmetry remains unbroken.  

\begin{figure}
\begin{center}
\leavevmode
\epsfysize=4in \epsfbox{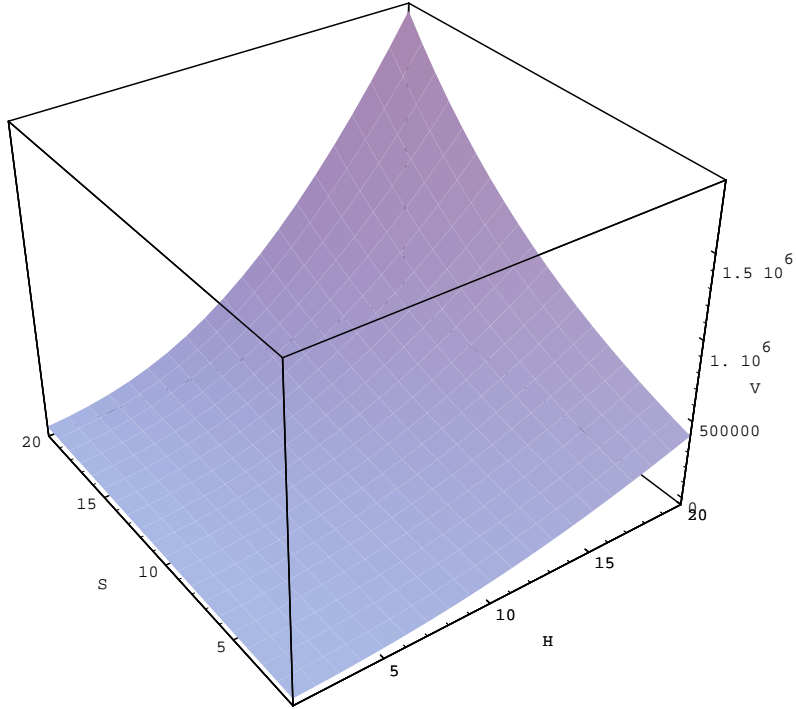}
\caption{The scalar potential $V(s,h)$. The graph corresponds 
to the example of $SU(3)$
as the gauge group, 
and assumimng $\langle  t\rangle=1$ for the internal modulus, 
$\langle  a \rangle=1.1$, $\lambda=0.1$, 
and $\lambda'=1$. The kinematically forbidden region 
of the $h$-$s$ plane has been 
excluded here; {\it i.e.,} the minimum of this plot is 
a point on the curve illustrated in Fig. 1, 
in this case $({\rm Re}\,s, h)=(2.66, 0.00044)$. }
\end{center}
\end{figure}
So far, the role of the 
intermediate scale has been masked. In the
following, we show that in the 
effective theory that we are 
considering, the free parameter $\mu$ 
in the nonperturbative superpotential
(25) is intimately related to the 
intermediate mass. Furthermore, we shall see
that the intermediate mass plays a role in allowing a sensible
hierarchy between the Planck scale 
and condensation scale, consistent
with the phenomenologically acceptable values of $\langle $Re$s \rangle$
and $\langle $Re$t\rangle$. For this, 
we shall give a rough argument below. Of course, the 
obvious effect that can immediately be associated with the intermediate 
mass is the shift it causes in the  
condensation scale, since in its 
presence the gauge coupling runs differently, as discussed in Section
3. 

\begin{figure}
\begin{center}
\leavevmode
\epsfysize=3.5in \epsfbox{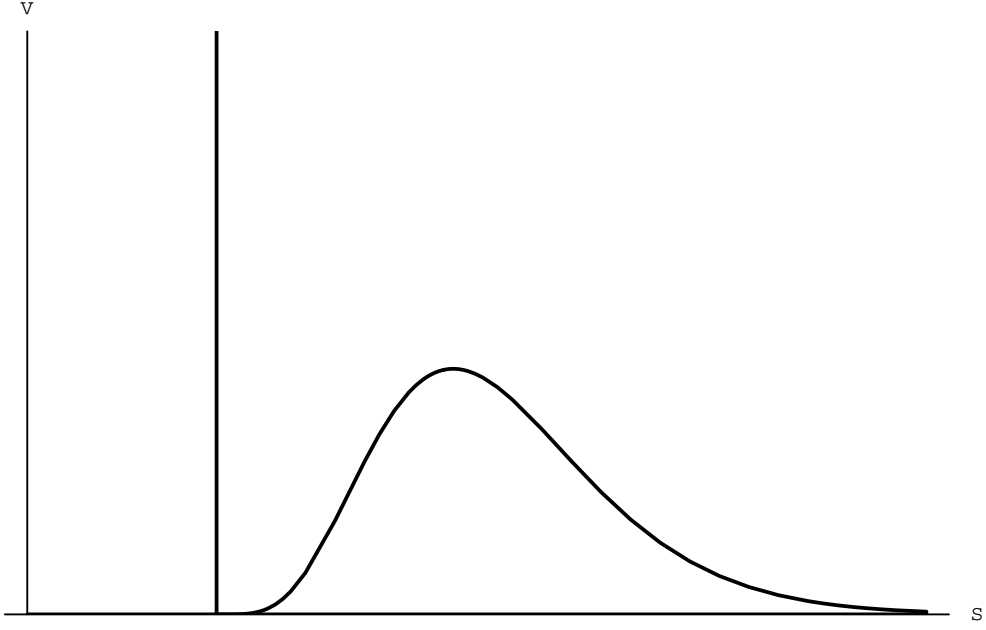}
\caption{The runaway behaviour of the dilaton in both directions.
In the  left direction the minimum 
corresponds to the effective coupling 
becoming strong. There, the potentiall `runs' into the kinematically 
forbidden region.} 
\end{center}
\end{figure}
In the presence $M_I$, the vacuum is characterized by two independent 
conditions:
\begin{equation} m=0, 
\hspace{0.3in} \langle W \rangle=0. \end{equation}
These two conditions together imply that:
\begin{equation} 
(t+\bar{t}-|a|^2)^{\widetilde{\Delta b}/2}\mu^{2b^<}
\widetilde{M}_I^{\Delta b}=1
. \end{equation}
Here, $\Delta b= b^>-b^<$ and $\widetilde{\Delta b}=b-b^<$. This 
can be re-written as follows:
\begin{equation}
s+\bar{s}-bG=b\left[
\frac{|\lambda a|}{3}\mu^{b^</\Delta b}\left(t+\bar{t}-
|a|^2\right)^{\widetilde{\Delta b}/2\Delta b} - 
1\right]^{-1}. \end{equation}
This equation should be viewed 
as a relation between $s$, $t$, and $\mu$
in the vacuum of the theory. 
Is this compatible with phenomenologically
acceptable values, $\langle  s+\bar{s} \rangle \sim 
{\cal O}(1)$ and $\langle  t+
\bar{t} \rangle 
\sim {\cal O}(1)$? If in eq. (22) we set $\langle  t+\bar{t} -
|a|^2\rangle\simeq 1$,\footnote{Notice 
that this does not restrict $M_I$ since 
$\lambda$ can be chosen small enough to give the assumed hierarchy
$M_I\ll M_{Pl}$.} which is also the assumption 
in the numerical analysis,
then it is easy to see that in order to get  
$\langle  s+\bar{s} \rangle \sim {\cal O}(1)$, 
\begin{equation}
\frac{\lambda \langle a \rangle}{3}\mu^{-|b^</\Delta b|} - 1 
\end{equation}
should be ${\cal O}(1)$. That is to say, 
for $b^</\Delta b$ of order unity,
\begin{equation} \mu \sim \lambda \langle a 
\rangle \sim {\cal O}(\widetilde{M_I}). \end{equation}
The free papramter of the effective superpotential for the
condensate  is, therefore, 
`locked' to the intermediate mass. 
This rough argument also shows that, with some fine tuning, it is at 
least possible in this scheme to obtain a phenomenologically
acceptable value for the dilaton, and at the same 
time achieve condensation and   
generate  the desired hierarchy.
\footnote{We hesitate to call this stabilization of the dilaton 
because the {\sl finite} value of $\langle  
Re\, S \rangle$ at which the potential runs to a minimum is at 
the boundary of the kinematically 
forbidden regime; $V$ is not smooth  there.} 
To see this,  consider eq. (21) again which together with eq.
 (24) tells us that:
\begin{equation} |\langle  h \rangle| \sim 
\exp\left(\frac{- \langle  s +\bar{s} \rangle}{6b^<} \right)\mu
\sim \exp\left(\frac{- \langle  s +\bar{s} \rangle}{6b^<}\right) 
\widetilde{M_I} .
\end{equation}
Again, we see that the parameters, which are admittedly model-dependent 
but are nevertheless, dictated by the presence 
of the intermediate mass and the choice of the gauge group 
can allow for a condensate whose vev is suppressed compared to 
the parameter $\mu$ which by requirement of phenomenology 
is of the order of the intermediate mass. 

\vspace{0.2in}

{\sc 5- Conclusions}

\vspace{0.3in}

Perhaps the most peculiar feature of the model of 
gaugino condensation that we have 
discussed above is the  running behaviour of the dilaton, 
which is schematically
shown in Fig. 3.  The {\sl finite} value of Re$ S$ that  
the potential ``runs" to is, as noted earlier,
on the boundary of the kinematically forbidden region,
and this  
value corresponds precisely to $1/g_{eff}(M_{cond}) \rightarrow 0$.  
We interpret this running of 
Re $S$ in both directions as a manifestation of 
S-duality which constrains the 
K\"ahler  potential which we have started with -- the
behaviour of the strong  and 
weak coupling (small and  large $S$, respectively)  regimes are 
alike. The intermediate scale serves basically to 
shift the renormalization running of the
gauge coupling and allow for a hierarchy between 
the unification and condensation scale
by shifting the condensation scale and/or the 
unification scale (see ref. 
\cite{unif}
for detailed discussion of the latter).
The intermediate mass (or rather, the vev of the gauge 
singlet) was assumed, but 
of course a realistic model should 
dynamically generate such an $M_I$ consistent
with phenomenology as discussed near 
the end of the previous section, and thereby 
giving a phenomenologically correct 
hierarchy of scales. This is of course a more
significant issue in the models where 
supersymmetry is broken by gaugino condensation
at the scale $M_{cond}$. 

However, as  we  have seen,  neither S-duality 
nor the 1-loop corrections to the 
dilaton in $K$ (including the 
dilaton dependent threshold corrections at $M_I$ are enough to 
break supersymmetry in such models.  
If one is to include any perturbative 
(1-loop) corrections to $K$, results such as 
those presented here or in 
ref. \cite{bgsdual} seem to indicate that it is more meaningful
to include the {\sl full} renormalization 
of the K\"ahler  potential, and all other terms that arise
at 1-loop in the supergravity and  
super-YM effective action which are relevant to gaugino 
condensation, such as 
\begin{equation} \int d^4\theta E  
(\frac{N_G \ln \Lambda ^2}{32 \pi^2}) (Re S) ^2 |W^{\alpha}W_{\alpha}|^2 . 
\end{equation} 
These have been recently calculated \cite{gjs}, 
and work along this direction is
under progress elsewhere \cite{ks}.  Indeed, as it has been argued by 
Banks and Dine,
if stabilization of the dilaton  
(and other moduli) and supersymmetry breaking
are really one and the same phenomena, as they appear to be, 
then stringy nonperturbative
corrections to K\"ahler  potential  are crucial and should be included 
\cite{bd}. A realization of
this proposal in the context
 of linear multiplet formulation of gaugino condensate appears in
ref \cite{bgw}. Of course, the exact form of these nonperturbative 
corrections are not 
yet understood. But one can perhaps expect that 
the recent developments in string dualities
can shed some light on the latter, and on the stabilization 
of string moduli and supersymmetry breaking.

\vspace{0.2in}

{\sc Acknowledgements} -  It is a pleasure to thank 
Mary K. Gaillard for inspiring this work 
and all the helpful discussions, and Yi-Yen Wu for 
collaboration on many parts 
of this work. This work was supported 
in part by the  Director, Office of Energy Research, 
Office of High Energy and Nuclear Physics, Division of High Energy
Physics of the U.S. Department of Energy under the Contract 
DE-AC03-76SF00098,
and by National Science Foundation under grant PHY-95-14797.

\newpage

% \vskip .3in
\appendix
\centerline{\large \sc Appendix}
\def\theequation{A.\arabic{equation}}

%       reset section commands

\catcode`\@=11

%\def\thesubsection{\Alph{subsection}.}
%\def\thesubsubsection{\arabic{subsubsection}.}
%
%\subsection{\sc The Role of the Gauge Coupling}
\setcounter{equation}{0}\indent%

{\sc S-Duality Transformations}
\setcounter{equation}{0}\indent

\vspace{0.3in}  

In this appendix, we review some elements of S-duality transformations
derived from the general formalism of ref. \cite{gz} (see also \cite{bgsdual}).
In the simplest case, in the presence of a YM field-strength $F_{\mu\nu}$,
the scalar fields parameterize the coset space $G/H$, where $G=SL(2,{\cal R})$,
is the (noncompact)  group of duality transformations and $H$ is its maximal 
compact subgroup $U(1)$.  Under the action of $SL(2,{\cal R})$, the bosonic 
component of the  dilaton transforms in the usual way:

\begin{equation} s\rightarrow s'=\frac{as-ib}{ics+d}  , \end{equation}
where \(a,b,c,d\) are real,  and $ad-bc=1$. The transformation of the
fermions is determined by the considering the invariance of the corresponding
kinetic terms and their coupling to the dilaton. One then obtains the
trasfomation property of the supermultiplet.  As shown in ref. \cite{bgsdual},
the transformation law (B.1) can be promoted to that of the dilaton (chiral)
supermultiplel as follows:
\begin{equation} S(\theta) 
\rightarrow \frac{aS(\theta ')-ib}{icS(\theta ') + d} = S'(\theta
')
,  \end{equation}
where
\begin{equation} \theta 
\rightarrow \theta '  =\left( \frac{ics+d}{-ic\bar{s} +d}\right)^{1/2}
\equiv \xi^{-1/2}\theta  ,  \end{equation}
and
\begin{equation} 
\psi_S \rightarrow \psi_S ' = \xi^{-1/2}(ics+d)^{-2}\psi_S . 
\end{equation}
Similarly, for the gaugino one finds:
 \begin{equation} 
\lambda_L \rightarrow \xi^{1/2}(ics+d)\lambda_L ,  \end{equation}
which implies that:
\begin{equation} W_{\alpha}(\theta)\rightarrow 
\xi^{1/2}(ics+d)W_{\alpha}(\theta ')  ;
\hspace{0.2in} U(\theta)\rightarrow\xi(ics+d)^2U(\theta ')  ,
\end{equation}
where $U$ is the composite field containing the gaugino condensate:
\( U=e^{K/2}H^3\). 
Here, $H$ is the usual chiral multiplet. Note that $U$ and
$H$ have different K\"ahler  
weights, therefore, $U$ differs from and ordinary
chiral superfield; in 
fact it can be shown to satisfy the constraint
$U=(\bar{\cal D}^2-8R)V$, where $V$ is a 
vector multiplet which contains the 
components of a linear multiplet 
and a chiral multiplet (\cite{bgsdual,bgt}).

It follows from the above 
transformation laws that the chiral field $H$
transforms as: 
\begin{equation} H\rightarrow(ics+d)^{1/3}H . 
\end{equation}
This, together with the fact 
that Re$S$ \( \rightarrow |icS+d|^{-2}\) Re$S$,
fixes (up to an S-invariant factor) 
the function $f(S,\bar{S})$ in the K\"ahler  potential
(21): $f=(S+\bar{S})^{1/3}$. Notice 
that the $T$-moduli are inert under S-duality
transformations.

\end{document}